\documentclass[abbrv,aps,prl,twocolumn,showpacs,preprintnumbers,amsmath,amssymb,nofootinbib]{revtex4}
\usepackage{amsmath,amssymb,amsfonts}
\usepackage{bm,graphicx,color}
\usepackage{graphics}
\usepackage{dcolumn} % Align table columns on decimal point
\usepackage{epsfig}

\begin{document}
\title{Excitonic condensation of strongly correlated electrons: the case of  Pr$_{0.5}$Ca$_{0.5}$CoO$_3$}
\author{Jan Kune\v{s}}
\email{kunes@fzu.cz}
\affiliation{Institute of Physics,
	Academy of Sciences of the Czech Republic, Cukrovarnick\'a 10,
162 53 Praha 6, Czech Republic}
\author{Pavel Augustinsk\'y}
\affiliation{Institute of Physics,
	Academy of Sciences of the Czech Republic, Cukrovarnick\'a 10,
162 53 Praha 6, Czech Republic}

\begin{abstract} 
We use a combination of dynamical mean-field model calculations and LDA+U material specific calculations
to investigate the low temperature phase transition in the compounds 
from the (Pr$_{1-y}$R$_y$)$_x$Ca$_{1-x}$CoO$_3$ (R=Nd, Sm, Eu, Gd, Tb, Y) family (PCCO).
The transition, marked by a sharp peak in the specific heat, leads to an exponential
increase of dc resistivity and a drop of the magnetic susceptibility, but no order parameter
has been identified yet. We show that condensation of spin-triplet, atomic-size excitons
provides a consistent explanation of the observed physics. In particular, it explains
the exchange splitting on the Pr sites and the simultaneous Pr valence transition. 
The excitonic condensation in PCCO is an example of a general behavior expected
in certain systems in the proximity of a spin-state transition.

\end{abstract}
\maketitle

The R$_x$A$_{1-x}$CoO$_3$ (R=La,..., and A=Ca, Sr, Ba) series
exhibits a variety of phenomena including thermally and doping driven spin-state crossover,
metal-insulator crossover, magnetic ordering or nanoscopic inhomogeneities.
The root cause of the rich physics are quasi-degenerate Co $3d$ atomic multiplets 
and their interaction with the crystal lattice or doped charge carriers.
The (Pr$_{1-y}$R$_y$)$_x$Ca$_{1-x}$CoO$_3$ (R=Nd, Sm, Eu, Gd, Tb, Y) family 
is unique among the cobaltites.  
A decade ago, Tsubouchi {\it et al}~\cite{tsubouchi02,tsubouchi04} observed a metal-insulator 
transition in Pr$_{0.5}$Ca$_{0.5}$CoO$_3$ associated with a 
drop of magnetic susceptibility and a sharp peak 
in the specific heat indicating the collective nature of the transition.
Subsequently, the transition was observed in other PCCO materials
with $x$ and $y$ in the ranges 0.2-0.5 and 0-0.3, respectively~\cite{fujita04,hejtmanek10,hardy13}. 
Despite the evidence for a continuous, or very weakly first 
order, phase transition and the experimental effort~\cite{hejtmanek13} 
no long-range order could be identified. The PCCO materials in this respect resemble
the much famous hidden order prototype URu$_2$Si$_2$~\cite{mydosh11}.
An important step towards understanding of the transition in PCCO was made by observation
of Pr$^{3+}\rightarrow$Pr$^{4+}$ valence transition
which take place simultaneously.
~\cite{knizek10,hejtmanek13}.
%rendering the Co $3d$ bands below $T_c$ 
%much closer to a stoichiometric filling than above $T_c$.
Another clue to the nature of the PCCO hidden order is
the exchange splitting of the Pr$^{4+}$ Kramers ground state
in the absence of ordered magnetic moments~\cite{hejtmanek10,knizek13,hejtmanek13}.
%Moreover, the shape of the Schottky peak indicates
%a spatially homogeneous exchange~\cite{hejtmanek13}.

The basic features to be captured by a theory of the PCCO hidden order 
are: \emph{i}) substantial increase of resistivity below $T_c$,
\emph{ii}) the sharp peak in the specific heat at $T_c$, \emph{iii}) the drop of the magnetic susceptibility
and the departure from the Curie-Weiss behavior of the Co moments below
$T_c$, \emph{iv}) the Pr valence transition,
\emph{v}) the exchange splitting of the Pr$^{4+}$ Kramers doublet in the absence
of ordered magnetic moments. 
More subtle effects include the increase of $T_c$ with pressure~\cite{fujita04},
the lattice response consisting primarily in reduction
of the Co-O-Co angle below $T_c$~\cite{fujita04}, and the apparent softness of the exchange field
on Pr and the lack of a clear x-ray signature of the spin-state transition~\cite{herrero12,hejtmanek13}. 

In this Letter we explain the physics of PCCO by formation of excitonic condensate (EC).
Motivated by observation of excitonic instability of correlated electrons close
to a spin-state transition~\cite{kunes14}, we have performed two types of investigations. 
First, we have studied the EC phase in a minimal model using the dynamical mean-field theory (DMFT)~\cite{dmft}
and calculated temperature ($T$) dependencies of the various physical quantities
across the transition.
Second, we have obtained a $T=0$ EC solution for PCCO using the density-functional 
LDA+U method~\cite{shick99,wien2k}.

Two-orbital Hubbard model (\ref{eq:hubbard}) captures the competition between the atomic high-spin (HS) 
and low-spin (LS) states and thus provides a minimal description of a solid with
a spin-state transition.
\begin{equation}
\label{eq:hubbard}
\begin{split}
&H=\frac{\Delta}{2}\sum_{i,\sigma}\left(n^a_{i\sigma}- n^b_{i\sigma}\right)+ 
\!\sum_{\langle ij\rangle,\sigma}\!\!\left(t_a a_{i\sigma}^{\dagger}a^{\phantom\dagger}_{j\sigma}+t_b
	b_{i\sigma}^{\dagger}b^{\phantom\dagger}_{i\sigma}\right) \\
%  &+  U\sum_i \bigl(n^a_{i\uparrow}n^a_{i\downarrow}+n^b_{i\uparrow}n^b_{i\downarrow}\bigr)+
%   (U-2J)\sum_{i\sigma} n^a_{i\sigma}n^b_{i-\sigma}\\
%   &+(U-3J)\sum_{i\sigma} n^a_{i\sigma}n^b_{i\sigma},
 &+  U\!\!\!\!\sum_{i,\alpha=a,b}\!\!\!n^{\alpha}_{i\uparrow}n^{\alpha}_{i\downarrow}
	+U'\sum_{i,\sigma} n^a_{i\sigma}n^b_{i\bar{\sigma}}
 +(U'-J)\sum_{i,\sigma} n^a_{i\sigma}n^b_{i\sigma}.
   \end{split}
\end{equation}
Here $a_{i\sigma}^{\dagger}$, $b_{i\sigma}^{\dagger}$
are the creation operators of fermions
with spin $\sigma=\uparrow,\downarrow$ in orbitals $a$ and $b$ on the site $i$ of a square lattice,
and $n^{a,b}_{i\sigma}$ are the corresponding occupation number operators.
The DMFT calculations using the impurity solver
of Werner~{\it et al.}~\cite{werner06} were performed for $U'=U-2J$, 
$U=4$, $J=1$, $t_a=0.4118$, $t_b=-0.1882$ and $\Delta=3.40$, assuming
eV to be the unit of energy. Details can be found in the Supplemental Material (SM).
The model captures the basic features of perovskite cobaltites:
nearly degenerate LS and HS atomic states, the energy scales of the bandwidths and
the interaction strength, a band gap/overlap being much smaller than the bandwidths,
and the dominant Co-Co nearest-neighbor hopping on a bipartite lattice preserving the orbital flavor.
The main approximation consists in neglecting the actual orbital degeneracy
of the $d$-shell. 

Linear response calculations~\cite{kunes14} predicted the model to exhibit
excitonic instability in the magnetic channel.
The spin-triplet EC order parameter for the model with 
SU(2) symmetric interaction is a vector
$\boldsymbol{\phi}_i=\sum_{\alpha\alpha'=\uparrow,\downarrow}\boldsymbol{\sigma}_{\alpha\alpha'}
\langle a^{\dagger}_{i\alpha}b^{\phantom\dagger}_{i\alpha'}\rangle$~\cite{kaneko12,halperin68b},
where $\boldsymbol{\sigma}$ are the Pauli matrices.
For the density-density interaction of Hamiltonian (\ref{eq:hubbard}) $\boldsymbol{\phi}$ is constrained to the $xy$-plane.
\begin{figure}
	\begin{center}
		\includegraphics[height=0.45\columnwidth,angle=270,clip]{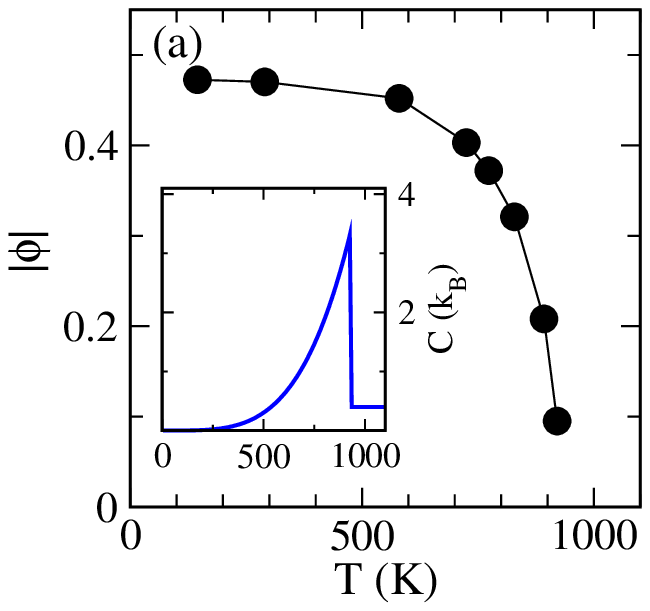}
		\includegraphics[height=0.51\columnwidth,angle=270,clip]{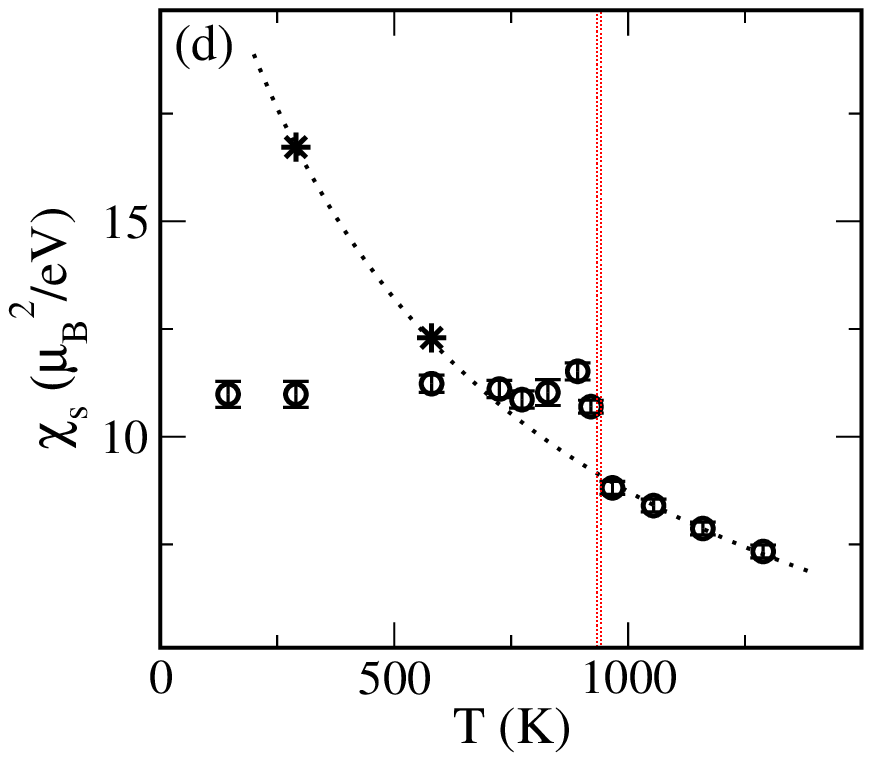}
                \includegraphics[height=0.45\columnwidth,angle=270,clip]{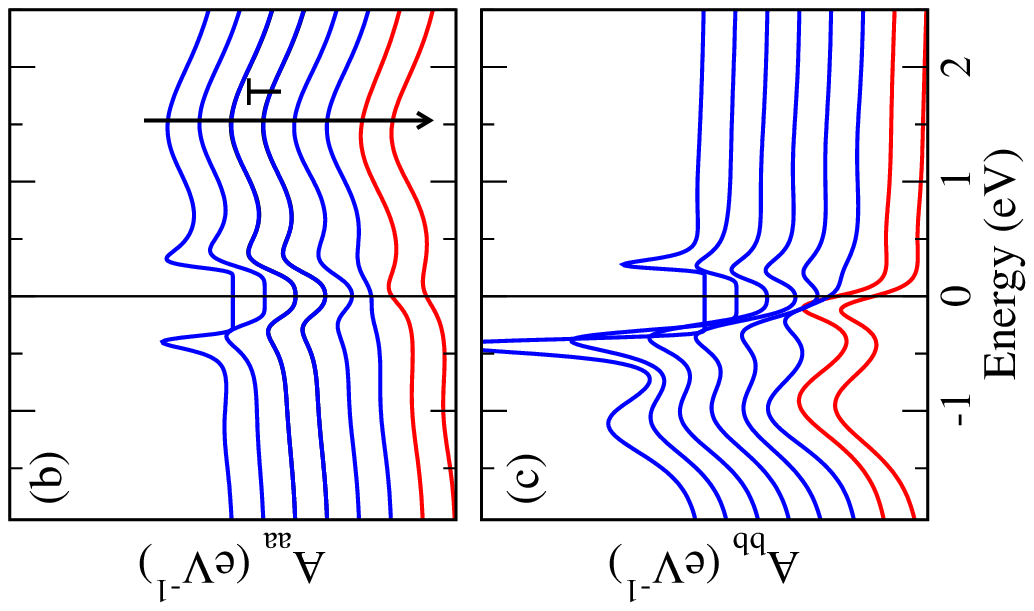}
                \includegraphics[height=0.51\columnwidth,angle=270,clip]{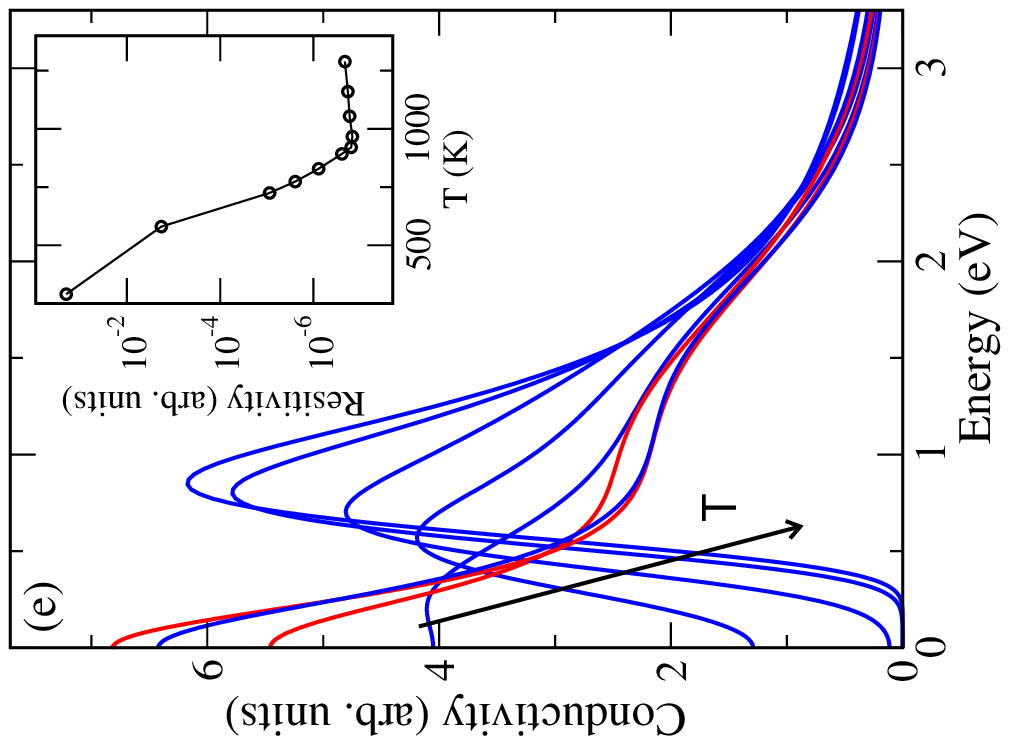}
	\end{center}
	\caption{\label{fig:ab_ch} The DMFT results for fixed particle density $n=2$.
         (a) The magnitude of the order parameter $|\phi(T)|$. The inset shows 
                specific heat $C(T)$. (b, c) The spectral functions $A_{aa}(\omega)$ and
                $A_{bb}(\omega)$ at $T=$1160, 968, 921, 892, 829, 725, 580, 290~K. 
                (blue curves for $T<T_c$, the red ones for $T>T_c$). The arrow marks 
                direction of increasing temperature.
                (e) The corresponding optical conductivity.
                The inset shows the dc resistivity.
                (d) The spin susceptibility $\chi_S(T)$ (circles with error bars) and
                $\chi_S(T)$ of the constrained normal phase solutions 
	        (dotted line).
	}
\end{figure}
First, we investigate the model at fixed particle density $n$ of 2 electrons per atom.
In Fig.~\ref{fig:ab_ch}a we show the evolution of the order parameter 
$\boldsymbol{\phi}$, which was chosen to point in the $x$-direction.
The inset shows the specific heat per atom with a typical mean-field shape. 
Non-zero $\phi_x$ is connected to appearance of a spin off-diagonal (anomalous) 
element of the self-energy (see SM for an example), which opens 
a gap in the one-particle spectra, Fig.~\ref{fig:ab_ch}b,c.
The gap opening is reflected in the behavior of the optical conductivity, Fig.~\ref{fig:ab_ch}e.
The Drude peak is suppressed below $T_c$ and  
the dc resistivity, shown in the inset, grows exponentially upon cooling.
While there are no ordered moments below $T_c$
the spin susceptibility $\chi_S(T)$, Fig.~\ref{fig:ab_ch}d, is strongly affected
by the EC transition. In the high-$T$ normal phase, thermally excited HS states lead
to Curie-Weiss $\chi_S(T)$. While HS states are present in the EC phase,
they are not free. The anomalous self-energy gives rise to an on-site 
hybridization between the LS and HS states which results in a $T$-independent Van Vleck $\chi_S(T)$.
The sign of the change of $\chi_S(T)$ at $T_c$ depends on details of the
system, in particular, a reduction of $T_c$ by doping, as discussed below,
leads to the same sign of $\chi_S(T)$ change as in the experiment.
The fact that HS population does not vanish in the EC phase
can explain the lack of changes in the x-ray spectra~\cite{hejtmanek13} typical
for the spin-state transition. 
%\begin{figure}
%	\begin{center}
%		\includegraphics[height=0.40\columnwidth,angle=270,clip]{aw.ps}
%		\includegraphics[height=0.49\columnwidth,angle=270,clip]{cond.ps}
%	\end{center}
%	\caption{\label{fig:aw_cond} The spectral functions $A_{aa}(\omega)$ (a) and 
%		$A_{bb}(\omega)$ (b) for $n=2$ at the temperatures $T=$1160, 968, 921, 892, 829, 
%		725, 580, 290~K. (c) The corresponding optical conductivity. The blue
%%	        curves correspond to $T<T_c$ while the red ones to $T>T_c$. 
%		Inset shows the $T$-dependence of the dc resistivity.}
%\end{figure}

The Co bands of PCCO differ from the above model in an important aspect. They are hole doped
in the normal state and their filling changes due to the Pr$^{3+}\rightarrow$Pr$^{4+}$ valence transition.
The isostructural valence transition points to a near degeneracy of the $f^2$ and $f^1$ 
states of the Pr $4f$ shell. The Pr ions therefore act as a charge reservoir providing 
electrons to the Co bands and can be modeled by fixing the chemical potential ($\mu$) 
in the above calculations.
\begin{figure}
        \begin{center}
                \includegraphics[height=0.40\columnwidth,angle=270,clip]{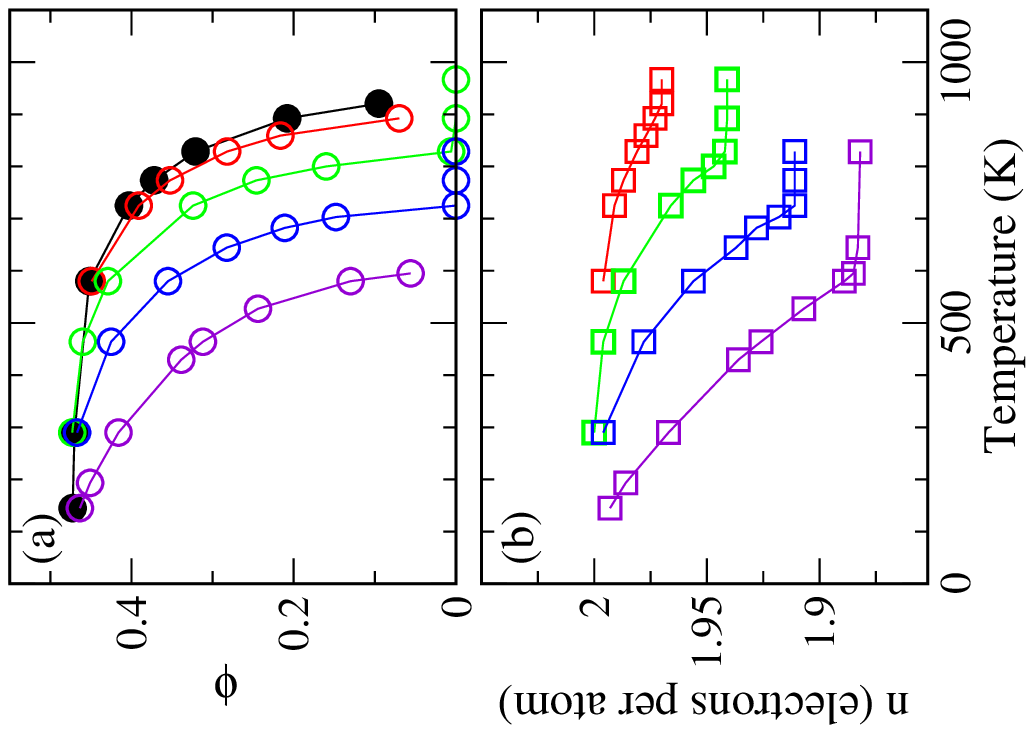}
                \includegraphics[height=0.49\columnwidth,angle=270,clip]{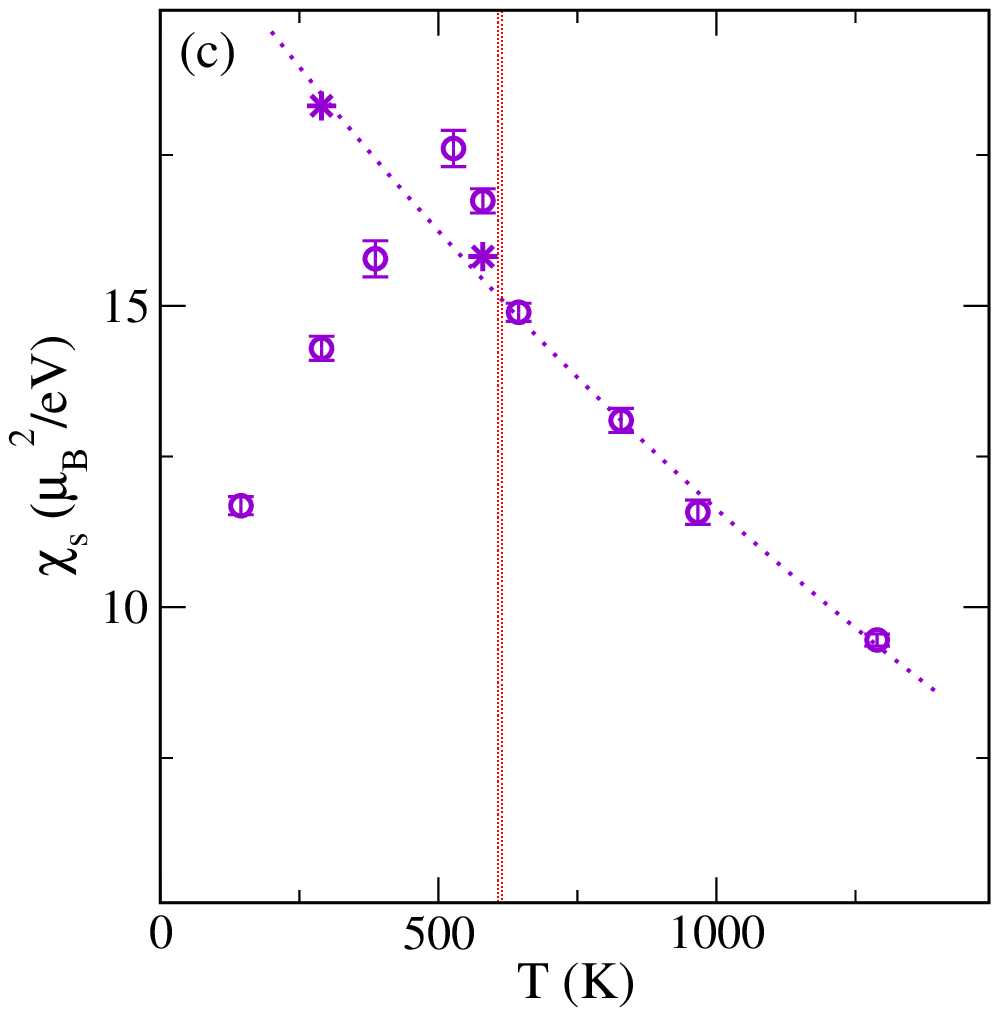}
        \end{center}
        \caption{\label{fig:ab_m} (a) The magnitude of the order parameter $|\phi(T)|$
                for various fixed chemical potentials (The $|\phi(T)|$ for $n=2$ 
                taken from Fig.~\ref{fig:ab_ch} is marked by black circles). 
                (b) The corresponding particle densities $n$ as functions of temperature. 
                The curves correspond to the doping of 0.03 (red), 0.06 (green), 0.09 (blue)
                and 0.12 (violet) holes per atom in the normal phase.
	        (c) The susceptibility $\chi_S(T)$ for the 0.12 hole doping. The symbols
                 have the same meaning as in Fig.~\ref{fig:ab_ch}.}
\end{figure}
In the following we present model results obtained with fixed $\mu$.
The particle density $n(T)$ in the normal phase is very weakly
$T$-dependent and thus can be used to label the different choices of $\mu$. 
In Fig.~\ref{fig:ab_m} we show $|\boldsymbol{\phi}|$ for dopings between 0.03 and 0.12 holes per atom.
Doping away from the half filling leads to a reduction of $T_c$. 
Unlike in the normal phase, $n(T)$ varies considerably below $T_c$.
With decreasing $T$ the system draws electrons from the reservoir, a process
controlled by competition between the condensation energy, favoring  
an equal number of $a$-electrons and $b$-holes, and the energy of adding electrons from the reservoir.
The present theory thus provides a simple connection between the Pr valence change
and the EC transition, and explains why these happen simultaneously~\cite{hejtmanek13}. 
The evolution of the one-particle spectra at fixed $\mu$ (see SM) 
is similar to Fig.~\ref{fig:ab_ch}b,c although the spectrum becomes fully gapped 
only at half filling. The behavior of the $\chi_S(T)$ for fixed $\mu$ 
is shown in Fig.~\ref{fig:ab_m}c.

The model calculations capture the features \emph{i-iv}. The \emph{\it i-iii} are generic
features of the EC transition in a half-filled system that
survive to a doped material kept at fixed $\mu$.
The feature \emph{iv} is accounted for by treating the Pr ions as a charge reservoir
for the Co bands. There are several limitations associated with the DMFT method as well as the model
itself. The mean-field character of the method is responsible for the extremely asymmetric
peak in the specific heat $C(T)$ as well as the kink in $n(T)$ at $T_c$. The experimental
$C(T)$ and $n(T)$~\cite{hejtmanek13} do not exhibit this pronounced asymmetry which can be 
explained by short-range EC correlations above $T_c$. 
The model also ignores the change of the lattice below $T_c$
consisting in bending of Co-O-Co without changing the Co-O bondlength. 
It enhances the $e_g$-$t_{2g}$ hopping, which provides a positive feedback to the EC transition. 
The transition with lattice taken into account is therefore expected to be sharper, 
perhaps even weakly first order, than in a purely electronic model. 

%\section{LDA+U study}
In order to test the EC scenario in a more realistic setting 
and to address the feature \emph{v}
we have performed a material specific calculation using LDA+U method. It 
roughly amounts to a $T=0$ static mean-field solution for Hamiltonian including all
electronic orbitals, the experimental crystal structure and unrestricted hopping.
Such calculation can answer the question whether the EC order in PCCO is plausible.  
The ability of the method to capture complex long-range orders
was demonstrated by Cricchio {\it et al.}~\cite{cricchio09,cricchio10}
on URu$_2$Si$_2$ and LaFeAsO.

%\begin{figure}
%	\includegraphics[width=\columnwidth,clip]{laco1.eps}
%	\caption{\label{fig:laco}
%		The rhombohedral unit cell of the cubic structure with the spin density depicted
%		as isosurfaces (positive--red, negative--blue) centered on the Co sites. The
%		O atoms are blue and La atoms are grey. A detail of the spin density distribution
%		for different EC solutions is shown on the right. 
%		We use isosurfaces of the collinear spin density
%		for the product solutions (a) and (b), and a tangent vector field
%		on the surface of vanishing normal component for the non-product solution (c).
%	}
%\end{figure}

%\subsection{Cubic structure}

Before presenting the results for the orthorhombic PCCO structure 
we discuss symmetry aspects of EC in a cubic crystal.
As in the model, the Hund's coupling 
selects the spin part of the order parameter to be a triplet. 
The orbital part describes a pair of an $e_g$-electron and a $t_{2g}$-hole, which
transforms as $E_g\times T_{2g}=T_{2g}+T_{1g}$ representation under the cubic symmetry operations.
General considerations suggest that only $T_{1g}$ pairs, $d_{xy}\otimes d_{x^2-y^2}$, 
$d_{xz}\otimes d_{x^2-z^2}$ and $d_{yz}\otimes d_{y^2-z^2}$ can condense.
The electrons and holes forming a $T_{1g}$ pair have large hopping
amplitudes along the same `in-plane' directions, while the electrons and holes forming
a $T_{2g}$ pair of the form $d_{xy}\otimes d_{z^2}$ maximize their hoppings in perpendicular directions,
which is detrimental to the condensation. Moreover, the electron-hole bonding
is stronger for a $T_{1g}$ than for a $T_{2g}$ pair.
The EC order in a cubic symmetry is thus characterized by nine parameters
$\phi^{\alpha}_{\beta}$, where $\alpha$ runs over the three Cartesian spin components and $\beta$
over three $T_{1g}$ orbital components. The anomalous part
of the Co $3d$ occupation matrices in terms of $\phi^{\alpha}_{\beta}$ can be found in SM.
We have verified that the numerical LDA+U solutions have these symmetry properties
by performing a series of calculations in cubic perovskite structure, which will
be reported separately.

\begin{figure}
\includegraphics[width=0.75\columnwidth,angle=270,clip]{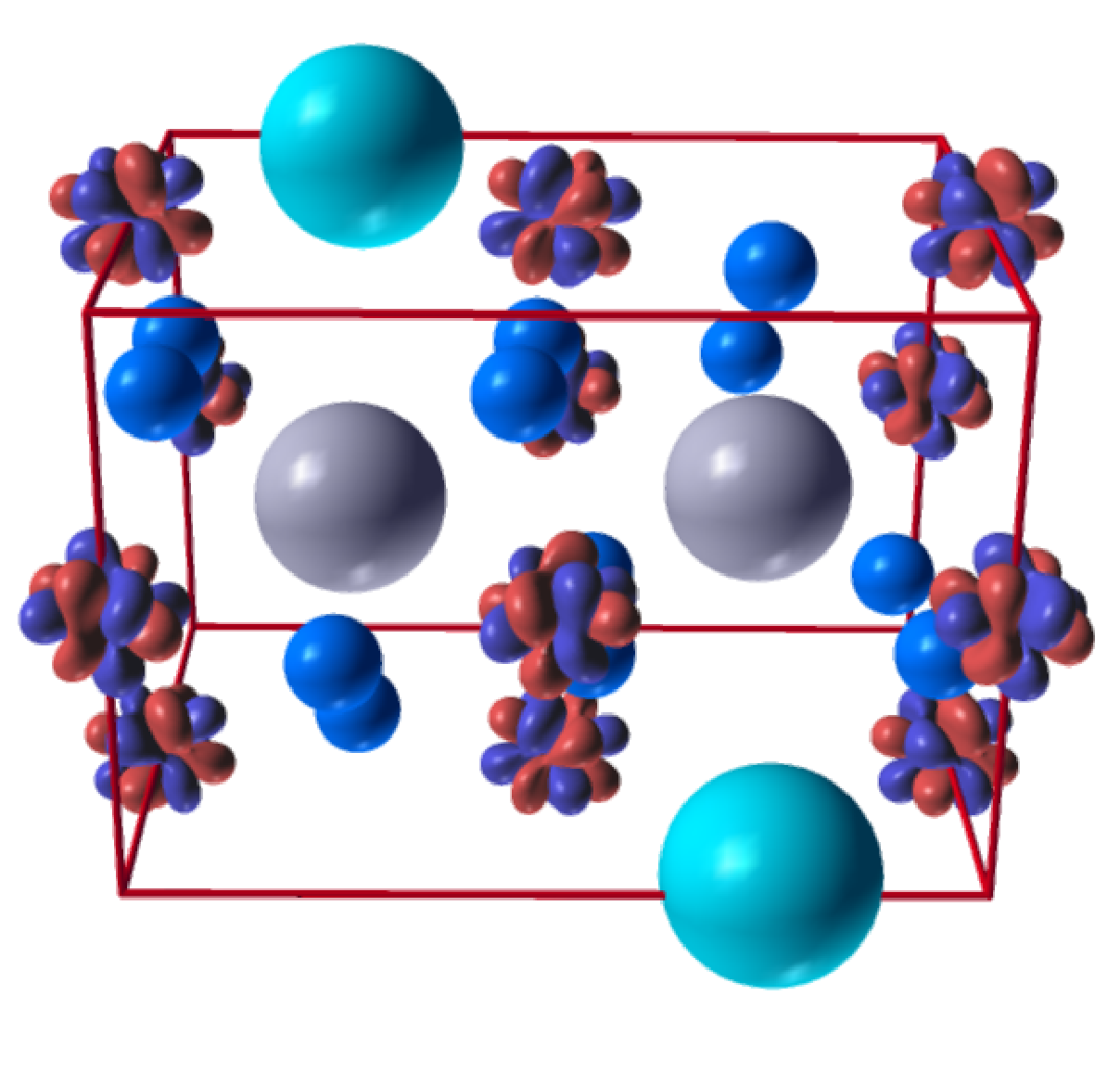}
\caption{\label{fig:prca1} The distribution of the collinear spin density (red and blue correspond to 
	positive and negative sign) around Co atoms in PCCO with O (blue), Ca (light blue) and Pr (grey). 
	}
\end{figure}

The LDA+U calculations for PCCO were performed in the structure of Ref.~\onlinecite{knizek10}
with a unit cell containing four Co sites.
On-site interaction parametrized with U=4~eV and J=1~eV was assumed for the Co $3d$ shells.
All Pr ions were assumed to be in the 4+ state, which was enforced by constraining the $f^1$ occupancy
in so called core treatment of the Pr $4f$ states. 
We found a stable EC solution with the total energy 230 meV per formula unit lower than the normal state one.
The EC was detected by appearance of spin-triplet terms in the Co-$3d$ occupation matrix. 
Reflecting the approximate cubic symmetry
of the Co sites, the orbital part of the anomalous terms is dominated
by the $T_{1g}$ components. The order parameter of the present solution
can be written as a product $\phi^{\alpha}_{\beta}=\varphi_{\beta}\otimes e_S^{\alpha}$
of a spin vector $e_S^{\alpha}$ pointing in arbitrary direction, but common to all Co sites, 
and an orbital pseudo-vector $\varphi_{\beta}$, shown in Table~\ref{tab:prca}. 
The product form of $\phi^{\alpha}_{\beta}$ with real elements results in the collinear 
spin-density distribution shown in Fig.~\ref{fig:prca1}. Inspection of $\varphi_{\beta}$
for symmetry related Co sites reveals an odd parity of the order parameter under the mirror
image $\sigma_h$ by a plane perpendicular to the $c$-axis. The EC solution does not exhibit
ordered local moments ($|\mathbf{m}|<0.03 \mu_B$ inside Wien2k atomic spheres).
The orbital resolved spectral functions can be found in SM.
\begin{table}
\caption{\label{tab:prca} The orbital parts of the EC order parameter for the four Co atoms
        the unit cell of PCCO in local coordinates tied to the CoO$_6$ octahedra.
        The sites 1-2 and 3-4 are connected by $\sigma_h$ symmetry.}
\centering
\begin{tabular}{ccccc}
\hline
\hline
     & 1 & 2 & 3 & 4  \\
\hline
    $\varphi_{yz}$ & 0.182 & 0.182 & 0.216 & 0.216  \\
    $\varphi_{zx}$ & 0.228 & 0.228 & -0.212 & -0.212  \\
    $\varphi_{xy}$ & -0.071 & 0.071 & -0.093 & 0.093 \\
\hline
\hline
\end{tabular}
\end{table}
\begin{figure}
\includegraphics[width=0.65\columnwidth,angle=270,clip]{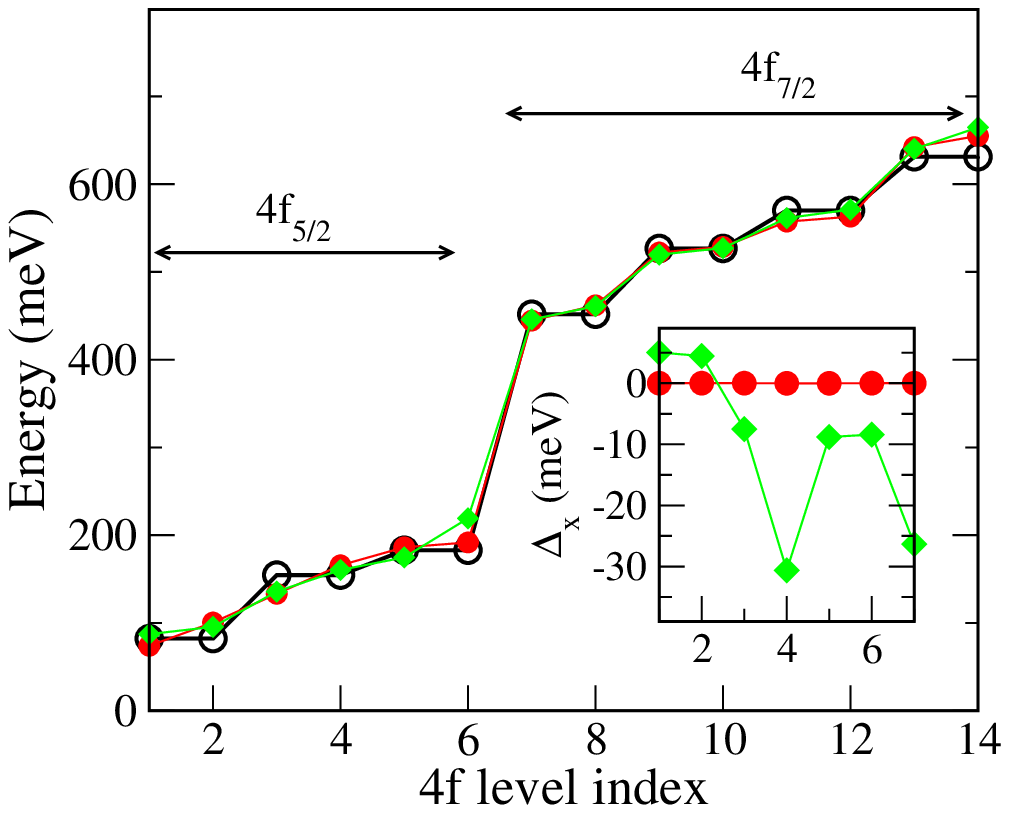}
\caption{\label{fig:prca} 
The spectrum of the Pr $4f$ states: 
no EC order (black), the self-consistent LDA+U solution with the $\sigma_h$-odd order parameter (red), 
with an artificial order parameter of the same magnitude containing a $\sigma_h$-even contribution
(green). The inset shows the exchange splitting of the $4f$ levels for the same order parameters
when spin-orbit coupling is not included. 
}
\end{figure}

Next, we address \emph{v}) the exchange splitting of the Kramers ground state
of the Pr$^{4+}$ ion. The EC with real $\phi^{\alpha}_{\beta}$ breaks 
the time reversal symmetry. However, we have to show that this symmetry breaking
is felt by the Pr moments. Microscopic analysis based on a multi-band Kondo impurity 
model can be found in SM. Here we use direct numerical calculation. 
To estimate the exchange splitting arising from the $4f$-ligand hybridization
we diagonalize the Kohn-Sham Hamiltonian of the EC solution with Pr $4f$ orbitals
included (with $E_{4f}$ inside the gap). This approach mimics the effect of 
the $f^1\rightarrow f^2\underline{L}$ virtual excitation~\cite{novak13}.
The calculated $4f$ spectrum is shown in Fig.~\ref{fig:prca}. The exchange splitting
induced by the EC order is clearly visible on top of the dominant spin-orbit 
and crystal-field splitting. The $4f$ spin-orbit coupling (SOC) is crucial. 
As shown in the inset of Fig.~\ref{fig:prca},
the EC order that is odd under the mirror image $\sigma_h$ does not couple to individual
$4f$ crystal-field states (without $4f$ SOC), which are either $\sigma_h$-odd or $\sigma_h$-even. 
It is only the SOC, which mixes the $\sigma_h$-odd and $\sigma_h$-even $4f$ functions and thus
allows the exchange splitting (see SM for more detail).
The exchange splitting of the order of 10~meV overestimates
the experimental values of a few Kelvin. This is not surprising given the approximations involved,
in particular the mean-field treatment of the Pr $4f$ shell which in reality presents
a complicated quantum impurity problem.

Spin-triplet excitonic condensation provides a comprehensive description
of the phase transition observed in the PCCO series. 
%While some of the points (i)-(iv) taken alone admit more conventional explanations,
%such as spin-state crossover, 
In particular, we are not aware of an alternative
theory of the exchange splitting of the Pr $4f$ states.
It is not clear at the moment why the excitonic condensation 
takes place in PCCO, but not in other cobaltites close 
to stoichiometric filling, e.g. LaCoO$_3$.
The answer is related to the nature of the lowest excited states
of the Co ion. The $S=2$ states tend to form a solid lattice on the 
LS background~\cite{knizek09,kunes11}, while $S=1$ states are
susceptible to the excitonic condensation. A phase separation is
another competing alternative in the doped systems.
 
The low-temperature phase of PCCO is an example of complex multipole order
which is detected only through its indirect effects.
Unlike URu$_2$Si$_2$ or LaOFeAs where the hidden order 
and nematicity arise from Fermi surface nesting~\cite{ikeda12,fernandes14}, PCCO are
strongly correlated oxides and the transition here is closer to condensation of 
preexisting composite bosons. The present mechanism of the excitonic condensation is quite
general and therefore it should be possible to find it in other
materials exhibiting singlet-triplet spin-state transitions~\cite{khaliullin13}.

We acknowledge numerous discussions with Z. Jir\'ak, P. Nov\'ak,
A. Kauch and D. Vollhardt.
The work was supported through the research unit FOR 1346 of the 
Deutsche Forschungsgemeinschaft and the grant 13-25251S of the
Grant Agency of the Czech Republic.

%\bibliography{prl3}

\end{document}